\documentclass[aps,prb,twocolumn,amsfonts,showpacs,longbibliography,superscriptaddress]{revtex4-2}
\usepackage{amsmath,amssymb,amsfonts,bm}
\usepackage{graphicx}
\graphicspath{{arXiv_1/}}
\usepackage{xcolor}
\usepackage{float}
\usepackage{placeins}
\usepackage{hyperref}
\hypersetup{
	colorlinks=true,
	linkcolor=blue,
	citecolor=blue,
	filecolor=black,
	urlcolor=blue,
}
\usepackage[capitalize]{cleveref}
\usepackage{enumitem}
\usepackage{booktabs}
\usepackage{array}
\usepackage{tikz}
\usetikzlibrary{arrows.meta,calc,decorations.pathreplacing,positioning}

\usepackage{algorithm}
\usepackage{algpseudocode}
\algnewcommand{\LineComment}[1]{\State \(\triangleright\) \textcolor{black}{\emph{#1}}}

\begin{document}
	
	\title{Statistical mechanics of classical fractons on a line}
	
	\author{Ylias Sadki}
	\email{ylias.sadki@physics.ox.ac.uk}
	\affiliation{Rudolf Peierls Centre for Theoretical Physics, University of Oxford, Oxford OX1 3PU, United Kingdom}
	\author{Abhishodh Prakash}
	\email{abhishodhprakash@hri.res.in}
	\altaffiliation{(he/him/his)}
	\affiliation{Harish-Chandra Research Institute (HRI), Prayagraj (Allahabad) 211019, India}
	\affiliation{Homi Bhabha National Institute (HBNI),  Mumbai 400094, India}
	
	\author{S. L. Sondhi}
	\email{shivaji.sondhi@physics.ox.ac.uk}
	\affiliation{Rudolf Peierls Centre for Theoretical Physics, University of Oxford, Oxford OX1 3PU, United Kingdom}

\begin{abstract}
We study the equilibrium statistical mechanics of one-dimensional classical Machian fractons: particles whose dynamics conserves a global dipole moment and whose Hamiltonian couples momentum differences through a position-dependent pair-inertia kernel. 
Compactly supported interaction kernels have a  divergence in the Gibbs partition function, and have been shown to break ergodicity and symmetry by forming non-equilibrium steady states with particle clusters, evading the Hohenberg-Mermin-Wagner-Coleman theorem. In this paper, we consider kernels with non-compact support and study their ergodic properties.
For exponentially decaying kernels, graph-Laplacian and matrix-tree bounds provide an extensive free energy suggesting that a putative statistical mechanical description is valid.
Similarly, uniform non-local kernels have a super-extensive free energy and require a Kac rescaling.
A generalized Hohenberg--Mermin--Wagner--Coleman argument, supported by finite-size scaling, implies symmetry-breaking density order parameter vanishes at all wave vectors melting the long-range translation-breaking density order of compact kernels.
To study the resulting equilibrium ensemble, we construct a nonreversible event-chain Monte Carlo (ECMC) algorithm that samples the coupled position-momentum phase space while preserving the dipole moment and total momentum. 
The ECMC sampling is shown to quantitatively match long time-averaged quantites in Hamiltonian dynamics. The equilibrium liquid exhibits preferred short-range clustering and strongly non-Gaussian single-particle momentum tails associated
with the correlated nature of positions and momenta. 
This paper provides a detailed investigation into the equilibrium liquid properties of the non-compact regime, whilst the companion paper investigates the mechanisms that relax the liquid.
\end{abstract}
	
	\maketitle
	
\section{Introduction}

The success of equilibrium statistical mechanics rests on a remarkable loss
of microscopic memory. Once the conserved quantities are fixed, a generic
many-body trajectory is expected to explore the allowed region of phase space,
so that late-time observables can be described by an ensemble rather than by
the details of the initial state. Ergodicity is the bridge between these
dynamical and statistical descriptions. There has been sustained interest in
understanding ergodicity breaking in both classical~\cite{Palmer1982BrokenErgodicity}
and quantum settings~\cite{NandkishoreHuse2015MBL,Abanin2019MBL}.
Common settings for such behavior involve quenched
disorder~\cite{Palmer1982BrokenErgodicity,Abanin2019MBL} or kinematic
constraints on the allowed dynamics~\cite{RitortSollich2003}.
More recently, the interplay of global symmetries with locality has provided
a robust route to ergodicity
breaking~\cite{Sala2020Fragmentation,Khemani2020Shattering}.
Classical fractons, whose study was recently
initiated~\cite{classical_fractons}, provide a simple Hamiltonian setting
in which this mechanism of ergodicity breaking can be
studied~\cite{machian_fractons}.

Fractons are particles or quasiparticles with restricted mobility~\cite{
	NandkishoreHermeleFractonsannurev-conmatphys-031218-013604,
	PretkoChenYou_2020fracton,GromovRadzihovsky2022fractonReview}. A particularly
transparent route to fractonic motion is the conservation of charge multipole
moments~\cite{Pretko_PhysRevB.95.115139}. If both charge and dipole moment are
conserved, an isolated particle cannot translate although suitable composites can remain mobile.
Here we study the classical continuum realization of this constraint:
identical particles on a line with translation invariance and a conserved
global dipole moment~\cite{classical_fractons}. Their kinetic energy couples
momentum differences through a position-dependent pair-inertia kernel.
For a kernel of compact support, generic many-body trajectories split into
dynamically disconnected clusters, producing global ergodicity breaking and
translation-breaking cluster order~\cite{classical_fractons,machian_fractons}.
Earlier simulations found persistent clustering over accessible times even
when the strict cutoff was replaced by an exponentially decaying
tail~\cite{classical_fractons,machian_fractons}. In this work and its
companion~\cite{apoorv}, we set out to determine whether the observed clusters
are true steady states or long-lived transients.

The focus of this paper is to establish the validity and applicability of
equilibrium statistical mechanics for fracton models with noncompact
pair-inertia kernels. After integrating out the momenta, we use
graph-Laplacian and matrix-tree bounds to establish a normalizable Gibbs
ensemble and its thermodynamic scaling (\cref{sec:free_energy}).
Exponentially decaying kernels, bounded above and below by exponentials with
the same rate, have an extensive free energy. Uniformly non-decaying kernels
with a size-independent positive lower bound instead require a Kac rescaling,
dividing the coupling by the particle number~\cite{KacUhlenbeckHemmer1963}.
We adapt the Hohenberg--Mermin--Wagner--Coleman argument to exclude long-range
density order, conditional on an equilibrium bound supported by finite-size
scaling (\cref{sec:mermin_wagner}). Short-range cluster correlations survive.

To study the resulting ensemble, we construct a nonreversible event-chain
Monte Carlo algorithm that samples the coupled position-momentum distribution
while preserving the dipole moment and total momentum
(\cref{sec:ecmc-algorithm}). The sampled equilibrium state is a liquid that
retains preferred short-distance clustering and has strongly non-Gaussian
single-particle momentum tails (\cref{sec:equilibrium_properties}). Its
structure factor agrees quantitatively with that obtained from Hamiltonian
dynamics.

In the companion paper~\cite{apoorv}, we perform the complementary dynamical
analysis by directly solving the many-body equations of motion for both
compact and noncompact kernels. There we find that ergodicity remains broken
for compact kernels and is restored for the noncompact kernels studied.
Individual particles exchange between clusters faster than the density
pattern itself relaxes; the slower structural relaxation connects the
dynamics to the equilibrium liquid found here.

\section{Fracton Hamiltonians and their ergodic properties}
\label{sec:fracton_hamiltonians}

We consider $N$ identical unit-charge particles with canonical coordinates
$(x_i,p_i)$ on a one-dimensional ring of circumference $L$, with separations understood	periodically. The thermodynamic limit is taken as $N,L\rightarrow\infty$ at
fixed density $\rho=N/L$. The two conserved global quantities relevant here
are the dipole moment $D=\sum_i x_i$ and total momentum $P=\sum_i p_i$; we
work in the sectors $D=P=0$.
The Machian fracton Hamiltonian is
\begin{equation}
	H = \frac{1}{2} \sum_{i>j}^N \left(p_i - p_j\right)^2 K\left(x_i - x_j\right),
	\label{eqn:main_hamiltonian}
\end{equation}
where $K(x)$ is taken to be an even, non-negative pair-inertia function. The dependence on both position and momentum differences makes the Hamiltonian invariant under uniform translations of either all positions or all momenta. The associated conserved quantities are the total momentum and the global dipole moment, respectively~\cite{classical_fractons}.

In contrast with
Newtonian mechanics, where inertia is an intrinsic property of each particle,
here it is relational: a particle moves only through its coupling to other
particles, in the spirit of Mach's relational view of
inertia~\cite{Mach1919ScienceOfMechanics}.  Related connections between
fractons and Mach's principle were discussed in
Ref.~\cite{Pretko_MachPhysRevD.96.024051}. We refer to this as Machian
dynamics.
The same symmetry-based program has been extended to continuum quantum
mechanics and to multipole-conserving models in phase space
\cite{sadki2025continuum,sadki2026phase}.  In dipole-conserving quantum
chains, it has also revealed a universal filling-driven transition between
strongly and weakly Hilbert-space-fragmented regimes
\cite{classenhowes2025universalfreezingtransitionsdipoleconserving}.

For the Hamiltonian in \cref{eqn:main_hamiltonian}, the canonical Gibbs partition function for indistinguishable particles at inverse temperature $\beta$ is~\cite{PathriaBeale2021}
\begin{equation}
	Z = \frac{1}{N!}\int \prod_{i} dx_i ~ \delta\left( \sum_i x_i \right) \int \prod_{i} dp_i ~ \delta\left( \sum_i p_i \right) e^{-\beta H}, \label{eq:partition_function_equation}
\end{equation}
where the delta functions fix the dipole and total-momentum sectors.

For a pair-inertia kernel of compact support, the Machian coupling switches
off exactly beyond a finite separation. Generic many-body trajectories then
split into dynamically disconnected clusters whose positions and membership
retain memory of the initial state.
Their late-time states are attractors in position-velocity space, rather than
in canonical phase space, and therefore do not violate Liouville's theorem.
Different initial conditions select different attractors, producing global
ergodicity breaking and translation-breaking cluster order even in low
dimensions~\cite{classical_fractons,machian_fractons}. 
In Ref.~\cite{machian_fractons}, partition-function arguments showed that separated clusters generate additional zero modes in the interaction-graph Laplacian, leading to a divergent Gibbs weight.
This provides a heuristic dynamical-selection principle for clustering, analogous to order by disorder~\cite{MoessnerChalkerOBDPhysRevLett.80.2929,ChalkerOBD2011}: configurations hosting more zero modes receive enhanced weight. Since the system is nonergodic, this is not an equilibrium explanation of clustering.
    
Subsequent work showed that nontrivial chaotic motion within the clusters
can coexist with this global nonergodicity, and that generic trajectories
exhibit a Janus point and a bidirectional arrow of time~\cite{babbar2025classical}. Related scale-invariant models exhibit cosmological fixed points together with
a bidirectional arrow of time~\cite{singh2026cosmological}.
	
The divergence of the Gibbs partition function for compact $K$ is established
in \cref{sec:free_energy}. The clustered state is therefore a nonequilibrium steady state, not an equilibrium
crystal~\cite{machian_fractons},
and its translation breaking does not contradict the equilibrium
Hohenberg--Mermin--Wagner--Coleman theorem~\cite{
	HohenbergPhysRev.158.383,MerminWagner_PhysRevLett.17.1133,
	Colemancmp/1103859034,mermin1967absence}.

Earlier simulations found persistent clustering over accessible times even when the strict cutoff was replaced by an exponentially decaying tail~\cite{classical_fractons,machian_fractons}.
Such a tail couples every pair at finite separation and lifts the exact zero modes into non-zero but possibly small soft modes.
Hence, clustered configurations can remain long-lived without permanently disconnected sectors~\cite{machian_fractons}.

We distinguish two noncompact cases by the large-separation limit
of the underlying infinite-line kernel, before it is periodized on the ring:
\begin{enumerate}
	\item \emph{Exponentially decaying $K(x)$}, for which $K(x)>0$ at every
	finite separation but
	$\lim_{|x|\rightarrow\infty}K(x)=0$.
	\item \emph{Uniformly non-decaying $K(x)$}, for which $K(x)$ does not vanish as
	$|x|\rightarrow\infty$. The bounds below treat the uniformly non-decaying
	case in which one fixed constant $\epsilon>0$ satisfies
	$K(x)\geq\epsilon$ at every separation.
\end{enumerate}
The distinction therefore concerns whether the coupling vanishes
asymptotically, not whether it is exactly zero at any finite separation.
\begin{figure}[!h]
	\centering
	\includegraphics[width=8.6cm]{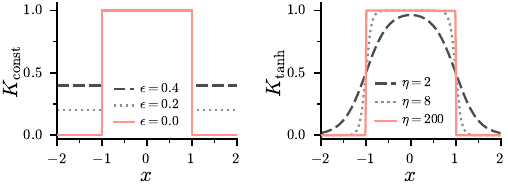}
	\caption{Representative examples of 
    (left) uniformly non-decaying $K(x)$, \cref{eq:K_box}, and (right) exponentially decaying $K(x)$, \cref{eq:K_tanh}. \label{fig:K}}
\end{figure}
For exponentially decaying $K(x)$, we use the concrete form
\begin{equation}
	K_{\tanh}(x) = \frac{1}{2}[\tanh(\eta(x+1)) - \tanh(\eta(x-1))],
	\label{eq:K_tanh}
\end{equation}
where $\eta^{-1}$ sets the width of the crossover. Increasing $\eta$ sharpens
the edge, and the limit $\eta\rightarrow\infty$ gives a compact top-hat
kernel as shown in \cref{fig:K}. On a finite ring this kernel has a strictly
positive minimum $\epsilon_L=K(L/2)$, but
$\epsilon_L\rightarrow0$ as $L\rightarrow\infty$. For uniformly non-decaying $K(x)$, we
use the form
\begin{equation}
	K_{\text{const}}(x) = \epsilon + (1-\epsilon) \left(\Theta(x+1) - \Theta(x-1)\right) 
	\label{eq:K_box}
\end{equation}
which approaches $\epsilon>0$ as $|x|\rightarrow\infty$.  Its lower bound is
independent of $L$, in contrast with $\epsilon_L$ for the exponentially
decaying kernel.

The methods and equilibrium properties for the different kernel classes are
summarized in \cref{tab:methods_summary}.

\newcommand{\yes}{\textcolor{green!60!black}{\checkmark}}
\newcommand{\no}{\textcolor{red}{$\times$}}
\newcommand{\notapp}{\textcolor{orange!80!black}{\footnotesize n/a}}
\newcommand{\rowlab}[1]{\parbox[t]{4.2cm}{\raggedright #1\strut}}
\newcommand{\colhead}[2]{\parbox[b]{2.95cm}{\centering #1\\[2pt]{\footnotesize #2}}}
\newcommand{\cell}[2]{\parbox[t]{2.95cm}{\centering #1\\[1pt]{\footnotesize\color{black!100}#2}\strut}}

\begin{table*}[t]
	\centering
	\setlength{\tabcolsep}{2pt}
	\renewcommand{\arraystretch}{1.15}
	\begin{tabular}{lcccc}
		\toprule
		& \multicolumn{2}{c}{\textbf{Noncompact} (infinite support)}
		& \multicolumn{2}{c}{\textbf{Compact} (finite support)}\\
		\cmidrule(lr){2-3}\cmidrule(lr){4-5}
		& \colhead{Decaying}{$K(x)\rightarrow0$ (e.g.\ tanh)}
		& \colhead{Uniformly nondecaying}{$K(x)\geq\epsilon>0$ (step)}
		& \colhead{Smooth}{differentiable cutoff}
		& \colhead{Step}{top hat}\\
		\midrule
		\addlinespace[2pt]
		\rowlab{ODE integration~\cite{machian_fractons,apoorv}} & \yes & \no & \yes & \no\\
		\rowlab{Event-driven MD~\cite{apoorv}} & \no & \yes & \no & \yes\\
		\rowlab{ECMC (this work)} & \yes & \yes & \no & \no\\
		\addlinespace[3pt]
		\midrule
		\addlinespace[3pt]
		\rowlab{$Z$ normalizable?}
		& \cell{\yes}{graph connected}
		& \cell{\yes}{graph connected}
		& \cell{\no}{cluster zero modes}
		& \cell{\no}{cluster zero modes}\\
		\addlinespace[3pt]
		\rowlab{$\log Z$ extensive as written?}
		& \cell{\yes}{$F=\mathcal{O}(N)$}
		& \cell{\no}{$F\sim N\log N$; Kac rescaling $K\rightarrow K/N$ restores extensivity}
		& \cell{\notapp}{$Z$ undefined}
		& \cell{\notapp}{$Z$ undefined}\\
		\addlinespace[3pt]
		\rowlab{Long-range density order forbidden?}
		& \cell{\yes}{subject to \cref{eqn:D_N_bound}; supported by numerical scaling}
		& \cell{\yes}{subject to \cref{eqn:D_N_bound}; supported by numerical scaling}
		& \cell{\notapp}{HMWC inapplicable}
		& \cell{\notapp}{HMWC inapplicable}\\
		\addlinespace[3pt]
		\rowlab{Short-range clustering}
		& \cell{\yes}{long-lived soft modes}
		& \cell{\yes}{long-lived soft modes}
		& \cell{\yes}{permanent sectors}
		& \cell{\yes}{permanent sectors}\\
		\addlinespace[2pt]
		\bottomrule
	\end{tabular}
	\caption{Applicability of the methods used in this paper to compact
		(finite-support) and non-compact (infinite-support) pair-inertia functions
		$K(x)$, the noncompact class being split into decaying and uniformly
		nondecaying kernels. The upper block collects the numerical methods and the
		lower block the equilibrium statistical mechanics. Whether ODE integration
		or event-driven molecular dynamics applies is set by the smoothness of
		$K(x)$ rather than by its support; these two rows follow the companion
		analysis in Ref.~\cite{apoorv}. The Hohenberg--Mermin--Wagner--Coleman
		(HMWC) row is conditional on the bound of \cref{eqn:D_N_bound} and is
		complemented by the finite-size scaling. For compact $K(x)$ the argument
		does not apply at all, since the partition function $Z$ is undefined. The
		remaining rows summarize
		\cref{sec:ecmc-algorithm,sec:mermin_wagner,sec:free_energy}.}
	\label{tab:methods_summary}
\end{table*}

\section{Free energy extensivity}
\label{sec:free_energy}
An equilibrium statistical-mechanical description requires more than formally
writing a Gibbs weight.  The partition function $Z$ must first be finite, so
that the Gibbs distribution can be normalized and equilibrium averages are
well defined.  Its large-system scaling must then be controlled: at fixed
density, the behavior of $F=-\beta^{-1}\log Z$ determines whether the free
energy per particle remains finite or whether the Hamiltonian requires an
additional size-dependent normalization. Thus, a divergent $Z$ rules out the
proposed Gibbs ensemble, extensive $F$ supports conventional thermodynamic
scaling, and superextensive $F$ identifies the normalization needed to recover
it. The following bounds establish these properties as
$N,L\rightarrow\infty$ at fixed density $\rho=N/L$.
\begin{enumerate}
	\item For compact $K(x)$, disconnected clusters generate zero modes~\cite{machian_fractons} so $Z = \infty$ and $F = -\infty$. Thus, the proposed Gibbs measure is not normalizable in the sector considered.
	\item For exponentially decaying $K(x)\sim e^{-\eta|x|}$, so that
	$K(x)\rightarrow0$ as $|x|\rightarrow\infty$, the free energy is
	extensive, $F=\mathcal O(N)$.
	\item For uniformly non-decaying $K(x)$ with $K(x)\geq\epsilon>0$ at every separation,
	$F\sim N\log N$: it scales \emph{superextensively}.
\end{enumerate}

The standard Gaussian integral over the $N-1$ independent momenta in \cref{eq:partition_function_equation} gives
\begin{equation}
	Z = \frac{1}{N!}\int \prod_i dx_i \delta(\sum_i x_i) \left( \frac{2\pi}{\beta} \right) ^{\frac{N-1}{2}} \sqrt{\frac{1}{\mathrm{det}'(L)}}, \label{eq:partition_main}
\end{equation}
where $L$ is the \emph{graph Laplacian} matrix, defined by 
\begin{equation}
	\frac{1}{2}\sum_{i < j} (p_i - p_j)^2 K(x_i - x_j) = \frac{1}{2}\sum_{i,j} p_i L_{ij} p_j.
\end{equation}
Here $\mathrm{det}'(L)$ denotes the product of eigenvalues of $L$, except for the trivial zero eigenvalue.  For standard background on graph Laplacians and spanning trees, see Ref.~\cite{Chung1997SpectralGraphTheory}.

In a previous work~\cite{machian_fractons}, graph Laplacian arguments provided a heurstic order-by-disorder argument for clustering.
For compact $K$, each disconnected cluster provides a zero eigenvalue (`zero mode') for $\mathrm{det}'(L)$, leading to a divergent weight in the partition function. Configurations with more clusters have more zero modes, suggesting a dynamical preference for these states.
However, for non-compact $K$, there are no exact zero modes in $\mathrm{det}'(L)$, yet nevertheless, small non-zero eigenvalues (`soft modes') still provide preference for clustered configurations.
	
\subsection{Exponentially decaying $K(x)$}
\label{sec:decaying-main}
Let us consider even, strictly positive kernels that vanish as
$|x|\rightarrow\infty$ and, more specifically, decay exponentially with
separation up to multiplicative constants independent of $N$ and $L$.  We
establish bounded linear scaling by obtaining complementary
lower and upper bounds on $F$.  On a ring of circumference $L$, define the
periodic distance
\begin{equation}
	r(x)=\min_{n\in\mathbb Z}|x+nL|,
\end{equation}
which is the length of the shorter arc between two particles and obeys
$0\leq r\leq L/2$.  We assume that there are constants $A,B,\kappa>0$, which
do not change as $N$ and $L$ grow, such that
\begin{equation}
	A e^{-\kappa r}\leq K(r)\leq B e^{-\kappa r}.
	\label{eq:exp-bounds-main}
\end{equation}
The exact kernel need not be a pure exponential: it only has to lie between
two exponentials with the same decay rate.  The tanh kernel in
\cref{eq:K_tanh}, used in Refs.~\cite{classical_fractons,machian_fractons},
satisfies this assumption explicitly.  For $r\geq0$,
\begin{align}
	K_{\tanh}(r)
	&=\frac{\sinh(2\eta)}{\cosh(2\eta r)+\cosh(2\eta)}\notag\\
	&=\frac{2\sinh(2\eta)e^{-2\eta r}}
	{1+e^{-4\eta r}+2\cosh(2\eta)e^{-2\eta r}}.
\end{align}
Because $0<e^{-2\eta r}\leq1$, the denominator in the second line lies
between $1$ and $2[1+\cosh(2\eta)]$.  Therefore
\begin{equation}
	\tanh(\eta)e^{-2\eta r}
	\leq K_{\tanh}(r)
	\leq 2\sinh(2\eta)e^{-2\eta r}.
	\label{eq:tanh-explicit-bounds-main}
\end{equation}
Thus \cref{eq:exp-bounds-main} holds for this kernel with
$\kappa=2\eta$, $A=\tanh\eta$, and $B=2\sinh(2\eta)$, all independent of
$N$ and $L$.

For each particle configuration, associate a weighted graph with one vertex
per particle and an edge of positive weight $K_{ij}=K(x_i-x_j)$ between
every pair $i,j$.  The weighted graph Laplacian is the matrix
\begin{equation}
	L_{ij}=
	\begin{cases}
		-K_{ij},&i\neq j,\\[2pt]
		\displaystyle\sum_{k\neq i}K_{ik},&i=j.
	\end{cases}
	\label{eq:laplacian-components-main}
\end{equation}
The diagonal entry $L_{ii}$ is therefore the total edge weight attached to
vertex $i$, also called its weighted degree.  Every row of $L$ sums to zero,
so the uniform vector $(1,1,\ldots,1)$ is a zero-eigenvalue vector.  This is
the already removed uniform-momentum mode.  Moreover, for any real vector
$\bm v$,
\begin{equation}
	\bm v^{T}L\bm v
	=\sum_{i<j}K_{ij}(v_i-v_j)^2\geq0.
	\label{eq:laplacian-positive-main}
\end{equation}
Thus no eigenvalue can be negative.  Because every $K_{ij}$ is strictly
positive, equality requires all $v_i$ to be equal, so the uniform vector is
the only zero mode.  All other eigenvalues are positive, and $\det'(L)$ in
\cref{eq:partition_main} means the product of these non-zero eigenvalues.

The factor $[\det'(L)]^{-1/2}$ in the integrand fixes the direction of both
arguments:
\begin{gather}
	\mathrm{det}'(L)\text{ lower}
	\ \Longrightarrow\ Z\text{ upper}
	\ \Longrightarrow\ F\text{ lower},
	\label{eq:bound-directions-lower}\\
	\mathrm{det}'(L)\text{ upper}
	\ \Longrightarrow\ Z\text{ lower}
	\ \Longrightarrow\ F\text{ upper}.
	\label{eq:bound-directions-upper}
\end{gather}
The second implication on each line reverses direction because
$F=-\beta^{-1}\log Z$.

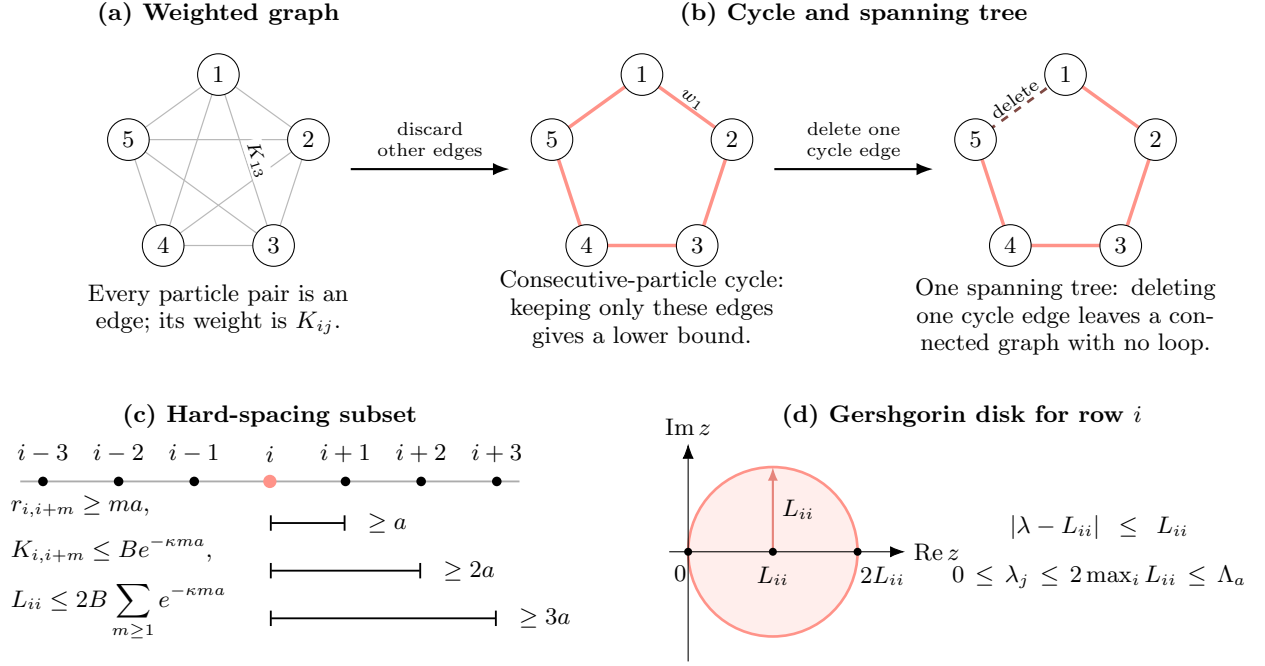
\begin{figure*}[t]
	\centering
    
    \definecolor{color_pink}{HTML}{FF9389}
	\begin{tikzpicture}[
		particle/.style={circle,draw=black,fill=white,minimum size=5.5mm,
			inner sep=0pt,font=\small},
		all edge/.style={draw=gray!55,line width=0.45pt},
		kept edge/.style={draw=color_pink!100!black,line width=1.3pt},
		removed edge/.style={draw=color_pink!50!black,dashed,line width=1.1pt},
		flow/.style={-{Latex[length=2.2mm]},line width=0.8pt},
		panel title/.style={font=\small\bfseries,align=center},
		panel note/.style={font=\small,align=center,text width=4.1cm}
		]
		\begin{scope}[shift={(0,0)}]
			\coordinate (a1) at (0,1.25);
			\coordinate (a2) at (1.19,0.39);
			\coordinate (a3) at (0.73,-1.01);
			\coordinate (a4) at (-0.73,-1.01);
			\coordinate (a5) at (-1.19,0.39);
			\foreach \i/\j in {1/2,1/3,1/4,1/5,2/3,2/4,2/5,3/4,3/5,4/5}
			\draw[all edge] (a\i)--(a\j);
			\path (a1)--node[pos=0.52,sloped,above,font=\scriptsize,
			fill=white,inner sep=1pt] {$K_{13}$} (a3);
			\foreach \i in {1,...,5}
			\node[particle] at (a\i) {$\i$};
			\node[panel title] at (0,2.05) {(a) Weighted graph};
			\node[panel note] at (0,-1.85)
			{Every particle pair is an edge; its weight is $K_{ij}$.};
		\end{scope}
		
		\begin{scope}[shift={(5.6,0)}]
			\coordinate (b1) at (0,1.25);
			\coordinate (b2) at (1.19,0.39);
			\coordinate (b3) at (0.73,-1.01);
			\coordinate (b4) at (-0.73,-1.01);
			\coordinate (b5) at (-1.19,0.39);
			\foreach \i/\j in {1/2,2/3,3/4,4/5,5/1}
			\draw[kept edge] (b\i)--(b\j);
			\path (b1)--node[pos=0.5,sloped,above,font=\scriptsize,
			fill=none,inner sep=1pt] {$w_1$} (b2);
			\foreach \i in {1,...,5}
			\node[particle] at (b\i) {$\i$};
			\node[panel note] at (0,-1.85)
			{Consecutive-particle cycle: keeping only these edges gives a lower bound.};
		\end{scope}
		
		\begin{scope}[shift={(11.2,0)}]
			\coordinate (c1) at (0,1.25);
			\coordinate (c2) at (1.19,0.39);
			\coordinate (c3) at (0.73,-1.01);
			\coordinate (c4) at (-0.73,-1.01);
			\coordinate (c5) at (-1.19,0.39);
			\foreach \i/\j in {1/2,2/3,3/4,4/5}
			\draw[kept edge] (c\i)--(c\j);
			\draw[removed edge] (c5)--node[pos=0.5,sloped,above,
			font=\scriptsize,fill=white,inner sep=1pt] {delete} (c1);
			\foreach \i in {1,...,5}
			\node[particle] at (c\i) {$\i$};
			\node[panel note] at (0,-1.95)
			{One spanning tree: deleting one cycle edge leaves a connected graph with no loop.};
		\end{scope}
		
		\node[panel title] at (8.4,2.05) {(b) Cycle and spanning tree};
		
		\draw[flow] (1.75,0)--(3.85,0)
		node[midway,above,font=\scriptsize,align=center]
		{discard\\other edges};
		\draw[flow] (7.35,0)--(9.45,0)
		node[midway,above,font=\scriptsize,align=center]
		{delete one\\cycle edge};
	\end{tikzpicture}
	
	\vspace{1.2em}
	\begin{tikzpicture}[
		font=\small,
		panel title/.style={font=\small\bfseries,align=center},
		panel math/.style={font=\small,align=center},
		measure/.style={<->,>=Latex,line width=0.7pt}
		]
		\begin{scope}[shift={(0,0)}]
			\node[panel title] at (0,2.35) {(c) Hard-spacing subset};
			\draw[gray!65,line width=0.7pt] (-3.3,1.48)--(3.3,1.48);
			\foreach \x/\lab in {-3/{i-3},-2/{i-2},-1/{i-1},1/{i+1},2/{i+2},3/{i+3}} {
				\fill[black] (\x,1.48) circle (1.8pt);
				\node[above=4pt] at (\x,1.48) {$\lab$};
			}
			\fill[color_pink] (0,1.48) circle (2.5pt);
			\node[above=4pt] at (0,1.48) {$i$};
			\draw[measure, |-|] (0,0.93)--(1,0.93)
			node[pos=1,right=5pt] {$\geq a$};
			\draw[measure, |-|] (0,0.30)--(2,0.30)
			node[pos=1,right=5pt] {$\geq2a$};
			\draw[measure, |-|] (0,-0.33)--(3,-0.33)
			node[pos=1,right=5pt] {$\geq3a$};
			\node[panel math] at (-2.0,0.36)
			{$\begin{aligned}
					&r_{i,i+m} \geq ma,\\[2pt]
					&K_{i,i+m} \leq Be^{-\kappa ma},\\[2pt]
					&L_{ii} \leq2B\sum_{m\geq1}e^{-\kappa ma}
				\end{aligned}$};
		\end{scope}
		
		\begin{scope}[shift={(7.8,0)}]
			\node[panel title] at (1.35,2.35) {(d) Gershgorin disk for row $i$};
			\fill[color_pink!16] (-1.15,0.55) circle[radius=1.12];
			\draw[color_pink!100!black,line width=1.0pt]
			(-1.15,0.55) circle[radius=1.12];
			\draw[-{Latex[length=2mm]}] (-2.52,0.55)--(0.6,0.55)
			node[right] {$\operatorname{Re}z$};
			\draw[-{Latex[length=2mm]}] (-2.27,-0.90)--(-2.27,1.98)
			node[above] {$\operatorname{Im}z$};
			\draw[-{Latex[length=1.8mm]},color_pink!90!black,line width=0.8pt]
			(-1.15,0.55)--(-1.15,1.67)
			node[midway,right=0pt,black] {$L_{ii}$};
			\foreach \x in {-2.27,-1.15,-0.03}
			\fill (\x,0.55) circle (1.5pt);
			\node[below left=2pt and 5pt] at (-2.,0.55) {$0$};
			\node[below=5pt] at (-1.15,0.65) {$L_{ii}$};
			\node[above right=4pt and 4pt] at (-0.25,-0.15) {$2L_{ii}$};
			\node[panel math,anchor=west,text width=4.6cm] at (0.75,0.55)
			{$\left|\lambda-L_{ii}\right|\leq L_{ii}$\\[8pt]
				$0\leq\lambda_j\leq2\max_iL_{ii}\leq\Lambda_a$};
		\end{scope}
	\end{tikzpicture}
	\caption{Graph constructions used for the free-energy bounds.  (a) The
		complete weighted graph has one vertex per particle and edge weight
		$K_{ij}$.  (b) Retaining consecutive-particle edges gives a cycle;
		deleting any one cycle edge produces a spanning tree, namely a connected
		graph with no closed loop.  Since all tree weights are non-negative,
		retaining only these $N$ trees lowers the matrix-tree sum.  (c) A minimum
		gap $a$ makes the distance to the $m$th neighbour grow at least as $ma$
		in both directions around the ring, so the corresponding edge weights
		form a convergent geometric series.  (d) For row $i$ of the graph
		Laplacian, the Gershgorin disk is centered at $L_{ii}$ and has radius
		$L_{ii}$.}
	\label{fig:matrix-tree-construction}
\end{figure*}

\paragraph{Lower bound on $F$: retain a cycle.}
For a fixed configuration, order the particles as one travels around the
ring.  Let $d_i$ be the clockwise gap from particle $i$ to particle $i+1$,
with particle $N+1$ identified with particle $1$.  These gaps satisfy
\begin{equation}
	d_i\geq0,
	\qquad \sum_{i=1}^N d_i=L.
	\label{eq:gaps-main}
\end{equation}
The lower bound follows from the \emph{weighted matrix-tree
	theorem}~\cite{matrix_tree_theorems,Chung1997SpectralGraphTheory}, which expresses the nonzero-eigenvalue
product of a positive weighted graph Laplacian as
\begin{equation}
	\mathrm{det}'(L)=N\sum_T\prod_{(ij)\in T}K_{ij}.
	\label{eq:matrix-tree-main}
\end{equation}
Here a spanning tree $T$ is a choice of $N-1$ edges that connects all
vertices without forming a closed loop; the weight of that tree is the
product of its edge weights.  The prefactor $N$ relates the product of the
$N-1$ nonzero eigenvalues to any cofactor of $L$.  Here a cofactor is the
determinant of the $(N-1)\times(N-1)$ matrix obtained by deleting one row and
the correspondingly numbered column.  The weighted-graph and spanning-tree
construction is shown in panels (a) and (b) of
\cref{fig:matrix-tree-construction}.

Every term on the right-hand side of \cref{eq:matrix-tree-main} is
non-negative.  We may therefore obtain a lower bound by keeping only the
cycle of edges joining consecutive particles.  Denote the weight of its
$i$th edge by $w_i$.  The shortest periodic distance across this edge is no
larger than $d_i$, and the lower bound in \cref{eq:exp-bounds-main} gives
\begin{equation}
	w_i\geq A e^{-\kappa d_i}.
\end{equation}
For the thermodynamic argument we may take $N\geq3$.  A cycle then has $N$
spanning trees, obtained by deleting any one of its $N$
edges.  If edge $i$ is deleted, the remaining product obeys
\begin{align}
	\prod_{j\neq i}w_j
	&\geq \left(A^{N-1}e^{-\kappa\sum_{j\neq i}d_j}=A^{N-1}e^{-\kappa(L-d_i)}\right) \notag\\
	&\geq A^{N-1}e^{-\kappa L}.
\end{align}
Summing these $N$ cycle trees and including the factor $N$ in
\cref{eq:matrix-tree-main} yields a bound that is independent of the
particular particle positions:
\begin{equation}
	{\mathrm{det}'(L)\geq N^2A^{N-1}e^{-\kappa L}.}
	\label{eq:det-lower-main}
\end{equation}

By \cref{eq:bound-directions-lower}, this bounds $Z$ from above.  The volume
of the translation-fixed configuration space is $L^{N-1}$, so
\begin{equation}
	Z\leq\frac{1}{N!}
	\left(\frac{2\pi}{\beta}\right)^{\frac{N-1}{2}}
	L^{N-1}N^{-1}A^{-\frac{N-1}{2}}e^{\kappa L/2}.
	\label{eq:Z-upper-main}
\end{equation}
At fixed density, $L=N/\rho$.  We use Stirling's large-$N$ formula
\begin{equation}
	\log N!=N\log N-N+\mathcal O(\log N).
	\label{eq:stirling-main}
\end{equation}
The
$N\log N$ contribution from $L^{N-1}$ cancels the one from $N!$.  Taking
$F=-\beta^{-1}\log Z$ then gives
\begin{equation}
	\frac{F}{N}\geq\frac{1}{\beta}
	\left[
	\log\rho-1+\frac{1}{2}\log\left(\frac{A\beta}{2\pi}\right)
	-\frac{\kappa}{2\rho}
	\right]
	+\mathcal O\left(\frac{\log N}{N}\right).
	\label{eq:free-energy-lower-main}
\end{equation}
Here $\mathcal O(\log N/N)$ means a correction bounded in magnitude by a
constant times $\log N/N$ at large $N$.
Thus $F/N$ cannot diverge to minus infinity.

\paragraph{Upper bound on $F$: use well-separated configurations.}
To obtain a lower bound on $Z$, we retain only configurations satisfying a
hard core constraint: every pair of neighbouring particles must be separated
by at least $a$,
\begin{equation}
	d_i\geq a,
	\qquad 0<a<\rho^{-1}.
	\label{eq:hard-gap-main}
\end{equation}
This constraint is imposed only to construct the bound; the full model still
allows smaller separations. Here $a$ is fixed as $N$ grows. The inequality on
$a$ ensures that $Na<L$,
so this subset has nonzero volume.  The two arcs from
particle $i$ to particle $i+m$ contain respectively $m$ and $N-m$ gaps.
The shorter periodic distance therefore satisfies
\begin{equation}
	r_{i,i+m}\geq a\min(m,N-m).
	\label{eq:periodic-neighbour-distance-main}
\end{equation}
Starting from particle $i$, count the first neighbour clockwise and the first
anticlockwise, then the second in each direction, and so on, until every
particle has been counted once. At step $m$, there are at most two particles,
each at least $ma$ away. Their combined contribution is therefore at most
$2B e^{-\kappa ma}$. Summing these bounds gives
\begin{align}
	L_{ii}=\sum_{j\neq i}K_{ij}
	&\leq\left( 2B\sum_{m=1}^{\infty}e^{-\kappa ma}=\frac{2B}{e^{\kappa a}-1} \right).
	\label{eq:degree-bound-main}
\end{align}
The factor two counts the two directions around the ring.  Extending the
finite sum to infinity only makes the right-hand side larger.  The last
equality is the geometric-series identity
$\sum_{m=1}^{\infty}q^m=q/(1-q)$ with $q=e^{-\kappa a}<1$.

Gershgorin's circle theorem~\cite{Varga2004Gersgorin} converts the degree bound into an eigenvalue
bound.  The theorem states that every eigenvalue of a matrix lies in at least one disk
centered at $L_{ii}$ and having radius
$\sum_{j\neq i}|L_{ij}|$.  For the Laplacian in
\cref{eq:laplacian-components-main}, that radius equals $L_{ii}$.  Since $L$
is symmetric and positive, its eigenvalues are real and non-negative; hence
every eigenvalue lies between $0$ and $2\max_iL_{ii}$.  Combining this fact
with \cref{eq:degree-bound-main},
\begin{equation}
	0\leq\lambda_j\leq\Lambda_a,
	\qquad
	\Lambda_a\equiv\frac{4B}{e^{\kappa a}-1}.
	\label{eq:eigenvalue-upper-main}
\end{equation}
The hard-spacing and Gershgorin constructions are shown in panels (c) and
(d) of \cref{fig:matrix-tree-construction}.

There are $N-1$ positive eigenvalues in $\det'(L)$, so
\cref{eq:eigenvalue-upper-main} implies
\begin{equation}
	{\mathrm{det}'(L)\leq\Lambda_a^{N-1}}
	\label{eq:det-upper-main}
\end{equation}
throughout the restricted region.

The volume of this region follows by writing $d_i=a+y_i$.  The
excess gaps satisfy $y_i\geq0$ and $\sum_i y_i=L-Na$.  For one fixed cyclic
ordering of the labels, the constraints on the $y_i$ define an
$(N-1)$-dimensional simplex of volume
$(L-Na)^{N-1}/(N-1)!$.  There are $(N-1)!$ cyclic orderings and an
overall translation factor $L$.  Hence the labeled hard-rod volume on the
ring is $L(L-Na)^{N-1}$.  Fixing the common translation removes the first
factor $L$, so the translation-fixed volume is
\begin{equation}
	[L-Na]^{N-1}=[L(1-a\rho)]^{N-1}.
\end{equation}
Using \cref{eq:det-upper-main} inside this restricted region gives
\begin{equation}
	Z\geq\frac{1}{N!}
	\left(\frac{2\pi}{\beta}\right)^{\frac{N-1}{2}}
	[L(1-a\rho)]^{N-1}\Lambda_a^{-\frac{N-1}{2}}.
	\label{eq:Z-lower-main}
\end{equation}
Using \cref{eq:stirling-main} gives the upper bound
\begin{multline}
	\frac{F}{N}\leq\frac{1}{\beta}\left[
	\log\left(\frac{\rho}{1-a\rho}\right)+\frac{1}{2}\log\left(
	\frac{4B\beta}{2\pi(e^{\kappa a}-1)}
	\right)-1\right]\\
	+\mathcal O\left(\frac{\log N}{N}\right).
	\label{eq:free-energy-upper-main}
\end{multline}
Thus $F/N$ cannot diverge to plus infinity.

\paragraph{Conclusion.}
The right-hand sides of
\cref{eq:free-energy-lower-main,eq:free-energy-upper-main} contain no
quantity that grows with $N$.
Therefore finite constants $c_-$ and $c_+$ exist such that
\begin{equation}
	c_-N+o(N)\leq F\leq c_+N+o(N).
\end{equation}
The notation $o(N)$ denotes sublinear terms, such as $\log N$.
Exponentially decaying $K(x)$ therefore produces no super-extensive term in the
free energy.  More precisely, these bounds show that $F/N$ remains bounded;
proving that it approaches a unique limit would require an additional
thermodynamic-limit argument~\cite{Ruelle1999RigorousResults}.

\subsection{Uniformly non-decaying $K(x)$}
\label{sec:infinite-main}

In the uniformly non-decaying case, $K(x)$ does not vanish as
$|x|\rightarrow\infty$.  For the bounds below we impose the stronger uniform
condition that there is a fixed constant $\epsilon>0$, independent of $N$
and $L$, such that
\begin{equation}
	\epsilon\leq K(x)\leq1
	\qquad\text{for every separation }x.
	\label{eq:infinite-kernel-bounds-main}
\end{equation}
The upper bound fixes the overall normalization of the kernel; the essential
condition is the size-independent positive lower bound.  The uniformly non-decaying 
kernel in \cref{eq:K_box} satisfies \cref{eq:infinite-kernel-bounds-main}
for $0<\epsilon\leq1$.

For any particle configuration, the graph introduced in
\cref{eq:laplacian-components-main} is complete because every pair weight is
positive.  Every spanning tree in \cref{eq:matrix-tree-main} contains exactly
$N-1$ edges, so \cref{eq:infinite-kernel-bounds-main} implies
\begin{equation}
	\epsilon^{N-1}
	\leq\prod_{(ij)\in T}K_{ij}\leq1.
	\label{eq:infinite-tree-weight-main}
\end{equation}
The number of these trees follows directly from
\cref{eq:matrix-tree-main}.  For the unit-weight complete graph,
$L=N I-\bm 1\bm 1^T$, so its $N-1$ nonzero eigenvalues are all equal to
$N$.  Equation~\eqref{eq:matrix-tree-main} then gives Cayley's formula~\cite{Cayley1889Trees}
\begin{equation}
	\lvert\mathcal T(K_N)\rvert=N^{N-2},
	\label{eq:cayley-main}
\end{equation}
where $K_N$ denotes the complete graph on $N$ labeled vertices and the
particle numbers distinguish those vertices.  Summing
\cref{eq:infinite-tree-weight-main} over the trees counted in
\cref{eq:cayley-main} gives
\begin{equation}
	\epsilon^{N-1}N^{N-2}
	\leq\sum_T\prod_{(ij)\in T}K_{ij}
	\leq N^{N-2}.
	\label{eq:infinite-tree-sum-main}
\end{equation}
Equation~\eqref{eq:matrix-tree-main} supplies one further factor of $N$.
Therefore
\begin{equation}
	(\epsilon N)^{N-1}
	\leq\mathrm{det}'(L)\leq N^{N-1}.
	\label{eq:infinite-det-bounds-main}
\end{equation}
The complete graph and a spanning-tree construction are illustrated in
panels (a) and (b) of \cref{fig:matrix-tree-construction}.

The bounds in \cref{eq:infinite-det-bounds-main} hold at every point of
configuration space.  Because the integrand in \cref{eq:partition_main}
contains $[\det'(L)]^{-1/2}$, the upper determinant bound gives a lower bound
on $Z$, whereas the lower determinant bound gives an upper bound.  Using the
translation-fixed volume $L^{N-1}$, define
\begin{equation}
	Z_{\mathcal K}\equiv
	\frac{L^{N-1}}{N!}
	\left(\frac{2\pi}{\beta}\right)^{\frac{N-1}{2}}
	(\mathcal K N)^{-\frac{N-1}{2}},
	\qquad \mathcal K\in\{\epsilon,1\}.
	\label{eq:infinite-ZK-main}
\end{equation}
The partition function is then bounded in the definite order
\begin{equation}
	Z_1\leq Z\leq Z_\epsilon.
	\label{eq:infinite-Z-bounds-main}
\end{equation}

At fixed density $L=N/\rho$, taking the logarithm of
\cref{eq:infinite-ZK-main} gives
\begin{align}
	\log Z_{\mathcal K}
	&=(N-1)\log L-\log N!
	+\frac{N-1}{2}\log\left(\frac{2\pi}{\beta}\right)\notag\\
	&\quad-\frac{N-1}{2}\log(\mathcal K N).
	\label{eq:infinite-log-ZK-main}
\end{align}
Let $F_{\mathcal K}=-\beta^{-1}\log Z_{\mathcal K}$.  Applying
\cref{eq:stirling-main} to \cref{eq:infinite-log-ZK-main} yields
\begin{multline}
	F_{\mathcal K}
	=\frac{N}{2\beta}\log N
	+\frac{N}{\beta}\left[
	\log\rho-1+\frac{1}{2}\log\left(
	\frac{\beta\mathcal K}{2\pi}\right)
	\right]\\
	+\mathcal O(\log N).
	\label{eq:infinite-FK-main}
\end{multline}
Since $F=-\beta^{-1}\log Z$, the partition-function inequalities in
\cref{eq:infinite-Z-bounds-main} reverse to give
\begin{equation}
	F_\epsilon\leq F\leq F_1.
	\label{eq:infinite-F-bounds-main}
\end{equation}
Both bounding expressions have the same leading term.  It follows that
\begin{equation}
	F=\frac{N}{2\beta}\log N+\mathcal O(N).
	\label{eq:infinite-free-energy-main}
\end{equation}
The free energy is therefore superextensive: $F/N$ grows logarithmically with
$N$.

The origin of this term is the determinant factor.  Pair couplings bounded
below independently of $N$ and $L$ make $\det'(L)$ scale as $N^{N-1}$ up to
factors exponential in $N$.  Its inverse square root contributes
$N^{-(N-1)/2}$ to the partition
function, whose logarithm produces the $N\log N$ term in
\cref{eq:infinite-free-energy-main}.  Size-dependent normalizations are
standard for long-range interacting systems~\cite{CampaDauxoisRuffo2009}.
Here the positive pair-inertia weights must be rescaled as
$K\rightarrow K/N$; in contrast, the random
couplings of the Sherrington--Kirkpatrick model use an $N^{-1/2}$ scaling
because their signs produce cancellations~\cite{SherringtonKirkpatrick1975}.
Under $K\rightarrow K/N$, the Laplacian is divided by $N$, so
$\det'(L)$ is divided by $N^{N-1}$ and the superextensive term is removed.
The Hamiltonian is also divided by $N$; Hamilton's equations therefore show
that the same trajectory is traversed on the rescaled time
$\tau\rightarrow N\tau$.

\section{Generalized Hohenberg--Mermin--Wagner--Coleman and the absence of long-range order}
\label{sec:mermin_wagner}

Having established that the Gibbs ensemble is normalizable for noncompact
$K$, and that its free energy is extensively (after 
rescaling where necessary), we can ask whether the resulting equilibrium free
energy can be minimized by a state that breaks continuous translation symmetry.  More
specifically, can long-range density order occur anywhere in the
finite-density, nonzero-temperature phase diagram?

   The Hohenberg--Mermin--Wagner--Coleman (HMWC) theorem is one of the central
no-go results of many-body physics: under its equilibrium and locality
assumptions, a continuous symmetry cannot be spontaneously broken in
sufficiently low dimensions~\cite{HohenbergPhysRev.158.383,
	MerminWagner_PhysRevLett.17.1133,Colemancmp/1103859034,
	mermin1967absence}.  It therefore provides
a sharp test of whether an apparently ordered one-dimensional state is stable under equilibrium statistical mechanics. 
   
   Our previous models with compact $K$ nevertheless exhibit late-time states
that break continuous translation symmetry and form a fracton
crystal~\cite{machian_fractons}.  This does not contradict HMWC: when the
interaction graph disconnects, additional momentum zero modes make the Gibbs
partition function divergent, so there is no normalizable equilibrium state
to which the theorem can be applied.  Noncompact $K$ removes this loophole.
Conditional on the equilibrium bound stated in \cref{eqn:D_N_bound}, the
generalized HMWC argument below shows that the density order parameter
vanishes.
Non-compact $K$ therefore forbids asymptotic long-range density order, although local structural correlations are still present.
This equilibrium conclusion agrees with the restoration of ergodicity in the
direct many-body dynamics of the companion paper~\cite{apoorv}.

We start with the classical Bogoliubov inequality used in the HMWC
argument, which is exact under the assumption of a Gibbs
measure~\cite{mermin1967absence}:
\begin{equation}
	\langle \lvert A \rvert^2 \rangle \geq \frac{k_B T \lvert \langle \{ A, C^*\} \rangle \rvert^2  }{\langle \{ C, \{ C^*, H \} \} \rangle },
\end{equation}
where $A$ and $C$ are any chosen phase-space observables, $H$ is the Hamiltonian, $C^*$ denotes complex conjugation, and $\{ \cdot, \cdot \}$ is the classical Poisson bracket~\cite{Arnold1989ClassicalMechanics}.
In order to prove the impossibility of long-range density order at wave-vector $G$, we chose 
\begin{align}
	A &= \rho_{G + k_L} = \sum_j e^{-i (G + k_L) x_j}, \\
	C &= C_{k_L} = \sum_j p_j e^{-i k_L x_j},
\end{align}
where $k_L = 2\pi/L$ is the smallest allowed wave-vector for a 1-dimensional system of length $L$. First, 
\begin{equation}
	\{ A, C^* \} = \sum_i \frac{\partial \rho_{G + k_L}}{\partial x_i} \frac{\partial C_{k_L}^* }{\partial p_i}
	= -i (G+k_L) \rho_G.
\end{equation}
Hence, the term in the numerator reduces to
\begin{equation}
	\lvert \langle \{ A, C^* \} \rangle \rvert^2 = (G + k_L)^2 \lvert \langle \rho_G \rangle \rvert^2.
\end{equation}
such that we now have
\begin{equation}
	\langle \lvert \rho_{G + k_L} \rvert^2 \rangle \geq \frac{k_B T (G + k_L)^2 \lvert \langle \rho_G \rangle \rvert^2}{D_N({k_L})},
\end{equation}
where we have now defined
\begin{equation}
	D_N(k_L) \equiv \langle \{ C_{k_L}, \{ C_{k_L}^*, H \} \} \rangle,
\end{equation}
which is the central quantity that must be bounded. On the left hand side, the triangle inequality gives
\begin{equation}
	\lvert \rho_{G+k_L} \rvert = \lvert \sum_j e^{-i (G + k_L) x_j} \rvert \leq N.
\end{equation}
Before defining an order parameter, we must distinguish a finite-size effect of
the fixed-dipole sector from genuine spontaneous symmetry breaking.  Under a
uniform translation $x_i\rightarrow x_i+a$, the density mode transforms as
\begin{equation*}
\rho_k\rightarrow e^{-ika}\rho_k.
\end{equation*}
In an unconstrained finite system, the Gibbs measure is invariant under every
$a$.  It would then follow that
$\langle\rho_k\rangle=e^{-ika}\langle\rho_k\rangle$ for every $a$, forcing
$\langle\rho_k\rangle=0$ whenever $k\neq0$.

\begin{figure}
    \centering
    \includegraphics[width=8.6cm]{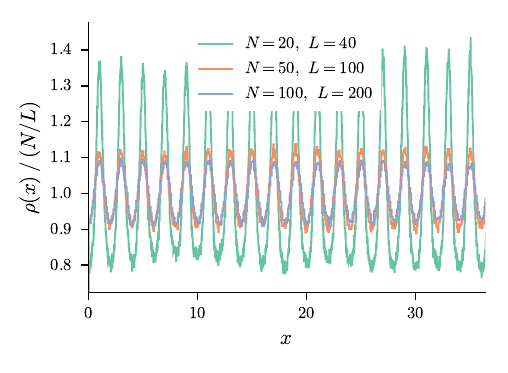}
    \caption{Fluctuation scaling of $\rho(x) = \sum_i \delta(x-x_i)$ obtained from binning ECMC data at $\rho = 0.5$, $\eta = 10$ and scaling $N=20,50,100$. Fluctuations are permitted at wavevector multiples of $2 \pi N / L$, or equivalently at real-space intervals of $L/N$. The fluctuations decrease with $L$ as required by the HMWC argument.}
    \label{fig:density_fluctuation_scaling}
\end{figure}

The fixed-dipole ensemble retains only a discrete part of this symmetry.  On
the ring, the position constraint in \cref{eq:partition_main} is understood as
the periodic delta function
\begin{equation*}
\delta_L(D)\equiv\sum_{n\in\mathbb Z}\delta(D-nL),
\qquad D=\sum_i x_i.
\end{equation*}
A uniform translation changes $D$ to $D+Na$.  It therefore preserves the
chosen dipole sector only when $Na$ is an integer multiple of $L$; the smallest
nonzero such translation is $a=L/N$.  Invariance under this residual
translation gives
\begin{equation*}
\langle\rho_k\rangle=e^{-ikL/N}\langle\rho_k\rangle.
\end{equation*}
The expectation value is consequently forced to vanish unless
$e^{-ikL/N}=1$.  Since the allowed ring wavevectors are $k=2\pi n/L$, a
nonzero first moment is symmetry-allowed only for
$k=2\pi \ell N/L$, with integer $\ell$.  These wavevectors correspond to
real-space periods $L/(\ell N)$.  Symmetry only permits these harmonics in the
constrained finite system; it does not require them or identify them as
spontaneous order.
The ECMC data in \cref{fig:density_fluctuation_scaling} show such oscillations
at finite size, with amplitudes that decrease as $L$ grows.

To test genuine spontaneous translation breaking, one must first select the
phase of a possible density wave and then take the limits in the appropriate
order, just as for the magnetization of a ferromagnet~\cite{WreszinskiZagrebnov2018Quasiaverages}. Introduce an
infinitesimal pinning field
\begin{equation}
H_h = H - h\sum_{i=1}^{N}\cos\left(G x_i + \varphi\right),
\end{equation}
where $\varphi$ fixes the origin of the modulation.  The order parameter is
then defined by
\begin{equation}
m_G \equiv \lim_{h \rightarrow 0^+} \lim_{\substack{N,L\to\infty \\ \rho \ \mathrm{fixed}}}
\frac{1}{N}\left\lvert\langle \rho_G \rangle_h\right\rvert.
\end{equation}

\begin{figure}[H]
	\centering
	\includegraphics[width=8.6cm]{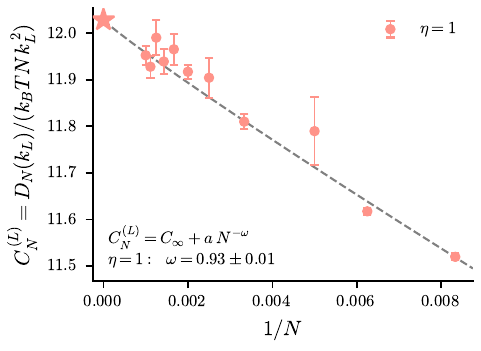}
	\caption{Plot of $C_N^{(L)}$ as defined in \cref{eqn:C_NL}, calculated from ECMC snapshots for the tanh kernel with $\eta=1$ and $\rho=1$. The dashed line is a fit to $C_N^{(L)}=C_\infty+aN^{-\omega}$. The fitted $C_\infty$ remains finite, supporting the bound in \cref{eqn:D_N_bound} over the simulated sizes. Under that bound, the argument excludes a nonzero density order parameter $m_G$.}
	\label{fig:D_N fit}
\end{figure}

The main inequality becomes
\begin{equation}
	m_G^2 \leq \frac{D_N(k_L)}{k_B T (G + k_L)^2}.
\end{equation}The central assumption (which we verify numerically in \cref{fig:D_N fit}) is the bound on $D_N(k_L)$. 
\begin{equation}
	D_N(k_L) \leq C k_B T N k_L^2 \ \forall \ N \ \mathrm{at~fixed~}\rho.
	\label{eqn:D_N_bound}
\end{equation}
where $C$ is an $N$-independent constant.
Under this assumption,
\begin{equation}
	m_G^2 \leq \frac{C N k_L^2}{(G + k_L)^2}.
\end{equation}
In one dimension, it is sufficient to consider a single $k = k_L$. (In two
dimensions, the standard proof instead sums over all $k$ and then takes a
continuum limit, with an additional ultraviolet bound on the schematic
$\sum_k \langle \lvert A_k \rvert^2 \rangle$ term
\cite{mermin1967absence}; we do not pursue that extension here).
With $k_L = 2\pi/L$ and $\rho=N/L=\mathrm{const}$, we have as $N \rightarrow \infty$:
\begin{align}
	&N k_L^2 = N \left( \frac{2 \pi} {L} \right)^2 = \frac{4\pi^2 \rho}{L} \rightarrow 0, \\
	&(G+k_L)^2 \rightarrow G^2 > 0.
\end{align}
Hence,
\begin{equation}
	m_G^2 \leq \frac{4\pi^2 \rho C}{L G^2} \rightarrow 0,
\end{equation}
so the symmetry-breaking density order parameter $m_G$ vanishes. The central assumption in \cref{eqn:D_N_bound} must therefore be tested.
We examine its large-$N$ behavior using ECMC snapshots, defining
\begin{equation}
	C_N^{(L)} = \frac{D_N(k_L)}{k_B T N k_L^2}.
	\label{eqn:C_NL}
\end{equation}
The data in \cref{fig:D_N fit} support this assumption for the simulated kernel, density, and system sizes.  Under this assumption, the argument above excludes a nonzero $m_G$ at any wave vector $G$.

The same argument also excludes a periodic array of clusters with varying populations.
Suppose there are $N_m$ particles at $X_m = m \frac{2\pi}{G}$, where $m$ is an integer labeling the cluster.
Then
\begin{equation}
	\rho_G = \sum_i e^{-i G x_i} = \sum_m N_m e^{- i G X_m} = \sum_m N_m = N.
\end{equation}
With the density-wave phase selected, this hypothetical state would therefore
have $m_G=1$, contradicting the earlier result $m_G=0$ under the bound in
\cref{eqn:D_N_bound}. Thus a periodic array of clusters with varying
populations cannot exist as a thermodynamic equilibrium state under this
assumption.

   With the conventional normalization, the static structure factor is
$S(k)=N^{-1}\langle\rho_k\rho_{-k}\rangle$~\cite{HansenMcDonald2013}; an ordered Bragg peak in this quantity scales as $\mathcal O(N)$ for a solid, whereas a liquid background remains $\mathcal O(1)$.
The first-moment $\langle \rho_G \rangle$ bound above does not by itself bound this second moment $S(k)$. We will devise a numerical check of this in the next section by finite scaling analysis.
Taken together, these two methods demonstrate the state is a genuine liquid, as opposed to a state whose density wave merely loses phase memory or translates.

	\section{Event-Chain Monte Carlo algorithm}
	\label{sec:ecmc-algorithm}
	
	Metropolis--Hastings Monte Carlo is a standard method for sampling probability distributions in statistical mechanics~\cite{Metropolis1953,Hastings1970}.
	Its transition kernel satisfies detailed balance, which makes the target distribution stationary; irreducibility and aperiodicity are additionally needed for convergence to that target~\cite{Tierney1994}.
	Reversibility is nevertheless not required for stationarity: lifted, nonreversible chains can instead satisfy global balance.
	Event-chain Monte Carlo (ECMC) uses this idea to produce directed motion through configuration space and samples rejection events directly; the original construction and a modern review are given in Refs.~\cite{BernardKrauthWilson2009,Krauth2021ECMC}.
	Because clustering dynamics in Machian-fracton models can be slow
	\cite{machian_fractons,babbar2025classical}, we implement ECMC to efficiently sample
	equilibrium distributions.
	
\begin{figure}[H]
	\centering
	\includegraphics[width=8.6cm]{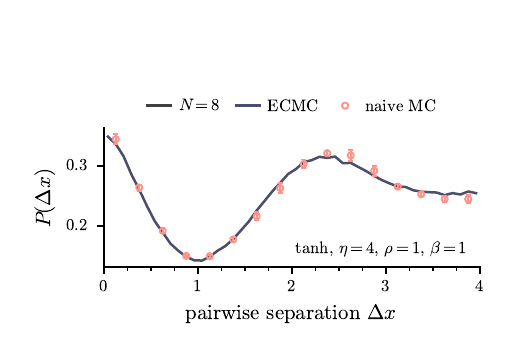}
	\caption{ECMC (\cref{alg:fracton_ecmc}) validated against the simple dipole-conserving MC algorithm (\cref{alg:naive_single_particle_mc_xp}) using the pairwise separation density for $N=8$ particles. }
	\label{fig:ECMC_validity}
\end{figure}

	For the exponentially decaying and uniformly non-decaying kernels considered here, we
	expect statistical mechanics to correctly measure equilibrium properties, as
	established in \cref{sec:free_energy}.
	The (unnormalized) Gibbs probability distribution over configurations $\{x_i, p_i\}$ is
	\begin{equation}
		P(\{x_i, p_i\}) = e^{-\beta H} \delta\left(\sum_i x_i\right) \delta\left(\sum_i p_i\right).
		\label{eq:prob_dist}
	\end{equation}
	It is best to directly sample $\{x_i, p_i\}$ instead of integrating out momenta, as each step involves only energy differences, which reduces to a small number of required computations, as the interactions are pairwise.
	The simplest algorithm would choose two random particles, move one to the left, and one to the right, naturally conserving the dipole moment.
	However, for the purposes of ECMC construction, it is best to use an algorithm with single particle moves, not two particle moves.
	Notice that a dipole conserving move may be proposed by choosing a random particle $i$ and displacement $\Delta_x$, then moving \emph{all} particles as follows
	\begin{equation}
			x_i \leftarrow x_i + \Delta_x ,\qquad
			x_{j \neq i} \leftarrow x_{j \neq i} - \frac{1}{N-1}\Delta_x
	\end{equation}
	Equivalently, we may move $x_i$ on its own, which individually will break dipole conservation, but then shift all particles using the translational invariance, to fix the dipole moment.
	This allows us to consider single particle moves, while still constructing a faithful algorithm that samples from $P(\{x_i, p_i\})$.
	We also consider momentum moves to explore the full phase space: 
	\begin{algorithm}[H]
		\caption{Single particle move Monte Carlo step}\label{alg:naive_single_particle_mc_xp}
		\begin{algorithmic}
			\State \textbf{input} $\beta$, $\{x_i, p_i\}_{i=1}^N$ 
			\State choose $i$ uniformly; draw $\Delta_x, \Delta_p \sim \mathrm{ran}(-1,1)$
			\State $\Delta E \gets H\!\left(\{x_i,p_i\}\ \mathrm{with}\ x_i{+}\Delta_x,\ p_i{+}\Delta_p\right) - H(\{x_i,p_i\})$
			\If{$\mathrm{ran}(0,1) < e^{-\beta \Delta E}$}
			\State $x_i \gets x_i + \Delta_x$; \quad $p_i \gets p_i + \Delta_p$
			\LineComment{Enforce dipole and total momentum conservation by translating:}
			\State $\{x_j\} \leftarrow \{ x_j - \frac{1}{N}\sum_k x_k\} $ 
			\State $\{p_j\} \leftarrow \{ p_j - \frac{1}{N}\sum_k p_k\} $ 
			\EndIf
			\State \textbf{output} updated $\{x_i, p_i\}_{i=1}^N$
		\end{algorithmic}
\emph{(Note: $A \leftarrow B$ denotes $A$ is updated to $B$).}
\end{algorithm}
This construction leaves $P(\{x_i,p_i\})$ stationary; subject to irreducibility, aperiodicity, and adequate equilibration, averages along the chain converge to equilibrium averages.
However, in this form, the algorithm must still run for very long times to obtain a representative sample of states from the distribution because clustering dynamics in Machian-fracton models can be slow.
For pair-factorizable Hamiltonians with well-defined event rates, the construction can be lifted to an \emph{event-chain} Monte Carlo algorithm~\cite{10.1063/1.4863991}.
Schematically, the road to a full ECMC construction for ${x_i, p_i}$ sampling is:
\begin{enumerate}
	\item Lift the process by introducing a single `chosen' particle that moves, which breaks reversibility.
	\item Take the limit of small step size, which makes the algorithm continuous.
	\item Analytically write a distribution for the time until the next rejection event.
	\item Finally, repeatedly sample the closest event, and pass the chosen index to the rejecting particle. 
\end{enumerate}
These lifting, continuous-time, and earliest-event steps follow the standard ECMC construction summarized above.
The full construction is detailed in Appendix~\ref{app:ecmc-construction}, which we summarize here:

\begin{algorithm}[H]
	\caption{Fracton Event-Chain Monte Carlo}\label{alg:fracton_ecmc}
	\begin{algorithmic}
		\State \textbf{input} $\beta$, $\{x_i, p_i\}_{i=1}^N$, sweep length $\mathcal{T}$, switch interval $\tau$, sampling interval $dt$
		\State $v \gets \mathrm{``position"}$; choose active particle $a$; $t \gets 0$
		\LineComment{$c_a$ denotes the active coordinate: $x_a$ in a position sweep, $p_a$ in a momentum sweep}
		\While{$t < \mathcal{T}$}
		\For{each $j \neq a$}
		\State draw $E_{aj}^{\mathrm{up}} \gets -\beta^{-1}\log(\mathrm{ran}(0,1))$
		\State invert $E_{aj}^{\mathrm{up}} = \int_0^{T_{aj}} \Delta V_{aj}(s)\,ds$ for rejection time $T_{aj}$
		\EndFor
		\State $T_{\mathrm{event}} \gets \min_{j \neq a} T_{aj}$; \quad $b \gets \arg\min_{j \neq a} T_{aj}$
		\LineComment{Record on the fixed grid $t = dt, 2dt, \dots$, not at events, to avoid sampling bias:}
		\State Record state with $c_a$ advanced to each grid time in $\left(t,\ \min(t + T_{\mathrm{event}},\, t_{\mathrm{switch}})\right)$
		\State $c_a \gets c_a + T_{\mathrm{event}}$; \quad $a \gets b$; \quad $t \gets t + T_{\mathrm{event}}$
		\If{$t$ has crossed a multiple of $\tau$}
		\LineComment{Swap sweep type}
		\State $v \gets \mathrm{``momentum"} \leftrightarrow \mathrm{``position"}$
		\LineComment{Re-enforce $\sum_i x_i = 0$ and $\sum_i p_i = 0$:}
		\State $\{x_j\} \gets \{x_j - \tfrac{1}{N}\sum_k x_k\}$
		\State $\{p_j\} \gets \{p_j - \tfrac{1}{N}\sum_k p_k\}$
		\EndIf
		\EndWhile
		\State \textbf{output} updated $\{x_i, p_i\}_{i=1}^N$
	\end{algorithmic}
\end{algorithm}
To verify the correctness of the ECMC approach, we compute $S(k) = N^{-1}\langle \rho_k \rho_{-k} \rangle$ for both the simple Monte Carlo single-step move, and the ECMC approach in \cref{fig:ECMC_validity}.

\section{Properties of the equilibrium distribution}
\label{sec:equilibrium_properties}
\begin{figure*}[t]
		\centering
		\begin{minipage}[t]{0.485\textwidth}
			\centering
			\textbf{(a)}\\[-0.5ex]
			\includegraphics[width=\linewidth]{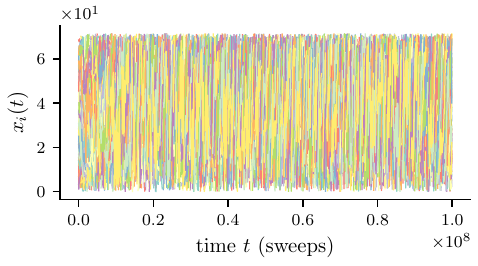}
		\end{minipage}\hfill
		\begin{minipage}[t]{0.485\textwidth}
			\centering
			\textbf{(b)}\\[-0.5ex]
			\includegraphics[width=\linewidth]{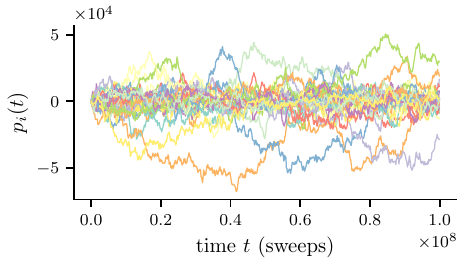}
		\end{minipage}
		
		\vspace{0.75em}
		
		\begin{minipage}[t]{0.32\textwidth}
			\centering
			\textbf{(c)}\\[-0.5ex]
			\includegraphics[width=\linewidth]{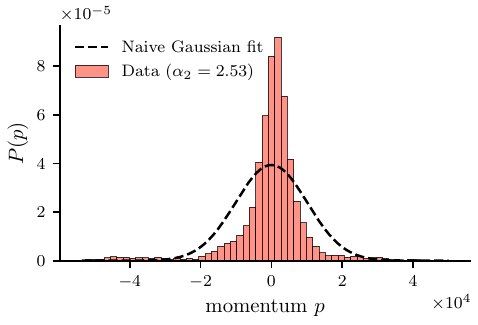}
		\end{minipage}\hfill
		\begin{minipage}[t]{0.32\textwidth}
			\centering
			\textbf{(d)}\\[-0.5ex]
			\includegraphics[width=\linewidth]{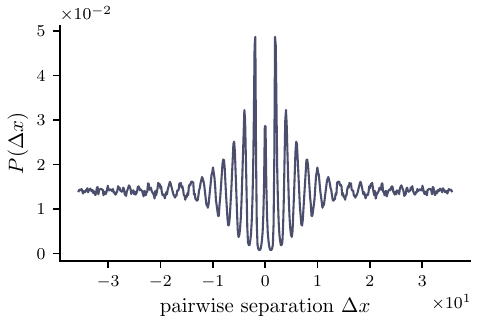}
		\end{minipage}\hfill
		\begin{minipage}[t]{0.32\textwidth}
			\centering
			\textbf{(e)}\\[-0.5ex]
			\includegraphics[width=\linewidth]{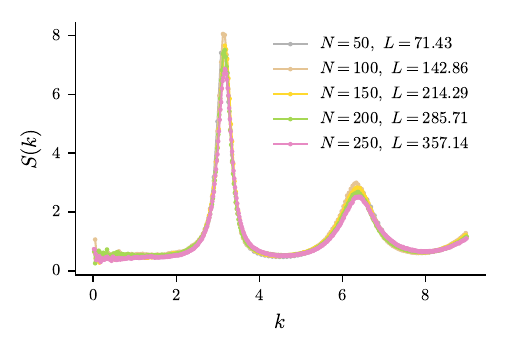}
		\end{minipage}
		\caption{
			Equilibrium data sampled by fracton ECMC for the tanh kernel with
			$N=50$, $\eta=10$, $\rho=0.7$, $\beta=0.5$.
			(a,b) Positions and momenta along the chain of samples; rare large-momentum
			events accompany rearrangements of the position configuration.
			(c) Single-particle momentum distribution $P(p)$, showing heavy non-Gaussian
			tails ($\alpha_2 \gtrsim 1$).
			(d) Pairwise separation distribution, showing preferred clustering length scales.
			(e) Static structure factor $S(k)$; the peaks are
			$\mathcal{O}(1)$, consistent with the absence of a Bragg peak. 
		}
		\label{fig:ecmc_equilibrium}
	\end{figure*}

Having established that for non-compact kernels, the equilibrium Gibbs state cannot be a long-range ordered crystal, we now comment on the nature of the resulting fracton liquid.
The ECMC algorithm allows us to directly probe properties of the equilibrium ensemble $e^{-\beta H}$, which we show in \cref{fig:ecmc_equilibrium}
In \cref{sec:mermin_wagner}, we demonstrated, using finite-size scaling of ECMC measurements, that long-range order and hence Bragg peaks are forbidden in the $N\rightarrow \infty$ limit.
This forbids Bragg peaks from appearing in the structure factor $S(k)$, which would scale as $\mathcal{O}(N)$ as for solids with crystalline order.
In \cref{fig:ecmc_equilibrium}(e), scaling $N$ in ECMC runs shows peaks that indeed do not scale with $N$, indicating conventional liquid correlations in structure.
    
For a conventional classical liquid with separable kinetic and potential energies, the Gibbs weight factorizes as $e^{-\beta H} = e^{-\beta H_{\mathrm{kin}}(\{p_i \})} e^{-\beta H_{\mathrm{pot}}(\{x_i \})}$.
The momenta are therefore distributed as Gaussian variables in the equilibrium ensemble.
As such, a Monte Carlo algorithm can independently sample the distribution $P(\{x_i\})$ using Metropolis-Hastings, and $P(\{p_i\})$ by sampling from the Gaussian distribution, to then obtain $P(\{x_i, p_i\}) = P(\{x_i\}) P(\{p_i\})$. This is, of course, not possible for fractons: positions and momenta are correlated by the non-separable Hamiltonian in \cref{eqn:main_hamiltonian}.
Intuitively, this is clear in the dynamics: the clustering in positions is manifested by large momentum differences between clusters.

In equilibrium states (see any time slice of \cref{fig:ecmc_equilibrium}(b)), the momentum of each particle is typically concentrated around a central value; yet, there are rare events where the momenta of a select few particles briefly becomes very large: see \cref{fig:ecmc_equilibrium}(a,b).
In dynamics, such an event induces a rearrangement of the clustering structure, for example by one particle jumping to a different cluster, while the density pattern remains broadly unchanged~\cite{apoorv}.
On a longer timescale, the density pattern itself melts.
	
These rare events push the individual momentum distributions $P(p_i)$ away from a Gaussian, introducing tails: \cref{fig:ecmc_equilibrium}(c).
We quantify this using the one-dimensional non-Gaussian parameter,
$\alpha_2=\langle p^4\rangle/[3\langle p^2\rangle^2]-1$~\cite{Rahman1964}.
A Gaussian has $\alpha_2=0$, whereas we find $\alpha_2\gtrsim1$ always.
The non-Gaussianity of momenta, even in equilibrium, highlights the correlated nature between positions and momenta.
For fractons with compact pair inertia functions~\cite{machian_fractons,babbar2025classical,apoorv}, the dynamics is strongly non-ergodic, with the momenta divergences driving clustering.
In the ergodic regime, momenta still plays an important role in dynamics~\cite{apoorv}, with large jumps in momenta generally coinciding with structural shifts in the positions of particles.
This translates in the equilibrium ensemble to a non-Gaussian distribution of momenta, with a small fraction of particles carrying large momentum.

\cref{fig:ecmc_equilibrium}(d) shows the probability distributions of pairwise separations, which is essentially the real-space version of the structure factor $S(k)$.
For the relatively low density system of $\rho = 0.7$ and sharp $\eta = 10$ shown in the figure, correlations between particles are observed over distances of $\sim 10$ particles, before decaying out.
This is characteristic of a liquid with short-range clustering, in direct contrast to the persistent long-range cluster order and dynamically disconnected sectors for the compact-support regime~\cite{machian_fractons}.

\begin{figure}[H]
	\centering
	\includegraphics[width=8.6cm]{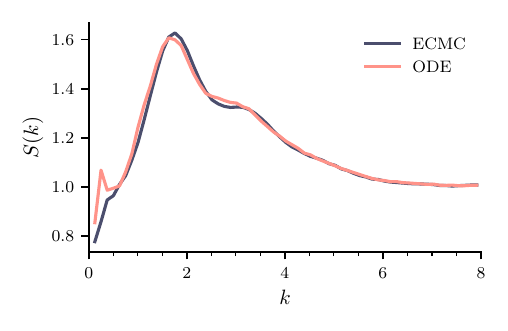}
	\caption{Plot of the structure factor $S(k)$ evaluated using both ECMC and ODE Hamiltonian time integration for $50$ particles, averaged over $10$ seeds. The kernel chosen is the tanh kernel with $\eta=3$. At this small $\eta$ the short range correlations are less structured, however higher $\eta$ equilibrates poorly with ODE integration.}
	\label{fig:ECMC_ODE_Match}
\end{figure}

Further, we confirm the ECMC approach does indeed sample the equilibrium reached by dynamics, by comparing ECMC to ODE integration in \cref{fig:ECMC_ODE_Match}. 
A strong match is quantitatively observed.

\section{Summary and future directions}

Earlier work on classical fractons established long-time clustering and ergodicity breaking in
many-particle systems with compactly supported pair-inertia
kernels~\cite{machian_fractons}, and persistent clustering in few-particle
systems with noncompact
kernels~\cite{classical_fractons,babbar2025classical}. Clustering with
exponential tails was also observed numerically in larger
systems~\cite{machian_fractons}. Taken together, these results suggested
that clustering is a robust feature of classical fractons. The present
work and its companion~\cite{apoorv} address the thermodynamic behavior
and relaxation of the many-body system with noncompact support,
distinguishing persistent density order from local and long-lived
clustering.

For the noncompact kernels considered here, we establish a normalizable
Gibbs ensemble and determine its thermodynamic scaling. Integrating out
the momenta expresses the configurational weight through the determinant
of an interaction-graph Laplacian. Graph-Laplacian and matrix-tree bounds
then show that the free energy is extensive for kernels bounded above
and below by exponentials with the same decay rate. For the uniformly
non-decaying kernels, the free energy instead contains the
superextensive term $N\log N/(2\beta)$, which is removed by the Kac
rescaling $K\rightarrow K/N$. These results establish the equilibrium
ensembles whose properties we investigate.

Within these ensembles, the generalized Hohenberg--Mermin--Wagner--Coleman
argument excludes a nonzero density-wave order parameter, subject to the
bound in \cref{eqn:D_N_bound}, which is supported by our numerical
finite-size scaling. The structure factor provides a complementary
test: its peaks do not grow proportionally to the particle number, as
required for Bragg order. Together, these results support a liquid
equilibrium state without long-range translation-breaking density order.
To sample this state, we construct an event-chain Monte Carlo algorithm
that preserves the dipole moment and total momentum while exploring the
coupled position-momentum distribution. The liquid retains pronounced
short-range clustering and strongly non-Gaussian single-particle momentum
tails, reflecting the correlations between positions and momenta.

The companion dynamical study~\cite{apoorv} finds persistent ergodicity
breaking for compact kernels and restoration of ergodicity for the
noncompact kernels examined. In the latter, melting proceeds in two
stages, distinguished by the correlations that relax. Tagged-particle
(self) correlations track the motion of individual particles: their
decay reflects particles leaving their original clusters and exchanging
between clusters. This exchange can occur while the collective density
pattern persists. Coherent density correlations instead track the
collective arrangement of particles: their slower decay marks the
relaxation of the cluster density pattern itself. It is this second,
structural stage that establishes the melting of density order and
connects the dynamics to the equilibrium liquid described here.
The quantitative agreement between structure factors obtained from
Hamiltonian evolution and ECMC sampling provides a direct test of
this connection.

The distinction between compact and noncompact support is also
reflected in the Laplacian spectrum. With compact support, separated
clusters become exactly disconnected and generate additional zero
modes, making the Gibbs partition function divergent. Noncompact tails
couple the clusters and lift these additional zero modes to nonzero
eigenvalues. For weak intercluster coupling, the former zero modes
become soft modes with small but finite eigenvalues. These soft modes
enhance the statistical weight of clustered configurations while
allowing a normalizable Gibbs ensemble. The equilibrium liquid can
therefore retain strong local clustering even though it has no
long-range density order.

An immediate open problem is to establish the bound in
\cref{eqn:D_N_bound} analytically and determine its range of validity
across kernels, densities, and temperatures. Extending the analysis to
higher dimensions would also determine whether noncompact classical
fractons can support an equilibrium ordered phase, and how such a phase
would be related to the nonequilibrium clustered states found for compact
support.

The dependence on the decay of the interaction kernel remains another
open question. A useful family is
\begin{equation}
    K_{\alpha}(r)=
    \left(1+\frac{\eta r}{\alpha}\right)^{-\alpha}.
\end{equation}
For every fixed $\alpha>0$, this kernel decays algebraically at large
$r$, while at fixed separation it approaches $e^{-\eta r}$ as
$\alpha\rightarrow\infty$ and the constant kernel as
$\alpha\rightarrow0^+$. The present exponential bounds do not determine
the thermodynamics of this family. Studying its Laplacian determinant
would establish the required size normalization and allow a comparison
of equilibrium correlations and relaxation across different decay
exponents. It would also clarify whether the
$\alpha\rightarrow0^+$ and thermodynamic limits commute.

Finally, the approach to compact support raises a dynamical question:
how do the particle-exchange and structural-relaxation times grow as
the tails are suppressed? Determining their dependence on tail strength,
density, and particle number would connect the few-particle clustering
results to the many-body liquid and clarify the order of the long-time,
thermodynamic, and compact-support limits. A quantitative relation
between these timescales, the small Laplacian eigenvalues, and the rare
large-momentum events would further connect the equilibrium statistics
to the mechanisms of melting.
\begin{acknowledgments}
	The authors would like to thank Werner Krauth and Gabriele Tartero for discussions on Event Chain Monte Carlo. S.L.S. acknowledges support from Leverhulme International Professorship grant LIP-202-014 and EPSRC grant EP\slash X030881\slash 1. 
	
	AI tools (Codex/ChatGPT/Claude) were used to assist in the ideas for proofs and prose, and for cleaning the LaTeX source.
	The authors reviewed and revised all AI-assisted changes and take full responsibility for the content of the manuscript.
\end{acknowledgments}

\appendix
\section{Event-chain Monte Carlo construction}
\label{app:ecmc-construction}

For fractons, we must sample both positions and momenta - this is at odds with the simplest ECMC, where the quadratic momenta factor out, such that $\pi(\{ x_i, p_i\}) \propto \pi(\{ x_i\}) \pi(\{ p_i \})$, so it is sufficient to sample positions, where interactions are pairwise.
For fractons, the algorithm must be generalized, but first we start with the standard lifted ECMC construction summarized in \cref{sec:ecmc-algorithm}~\cite{10.1063/1.4863991,Krauth2021ECMC}.
Consider an energy of configurations
\begin{equation}
	E = \sum\limits_{i < j = 1}^N V(x_i - x_j).
\end{equation}
Where we then want to sample from
\begin{equation}
	\pi(\{x_i \}) = e^{-\beta H( \{ x_i \}) }.
\end{equation}
The Metropolis--Hastings algorithm would proceed by proposing random moves, first selecting a random particle and a random displacement, and then accepting the move with a probability depending on the energy change.
En route to the ECMC construction, we first ``lift'' our process, then make the algorithm continuous.
To lift the process, we introduce a tagged particle, which we propose moves for, whilst not moving other particles.
This tagged particle is always moved in one direction, e.g. the positive $x$ direction.
The particle moves forward until one of the pair interactions rejects the move, then the tagged particle label is passed on. The direction of motion remains the same.
The directed lifting suppresses diffusive backtracking: its moves are closer
to ballistic motion than to a random walk~\cite{10.1119/5.0176853}.

Correctness of the lifted process is established through global balance.
We posit that the equilibrium distribution does not depend on the tagged particle, so
\begin{equation}
	\pi(\{ x_i \}, a, v) = \frac{1}{N}\pi(\{x_i\}),
\end{equation}
where $a$ labels the tagged particle index,

The process is as follows.
Particle $a$ is considered for a fixed displacement $\Delta x$.
The pairwise interactions are considered individually to accept or reject the move one at a time.
$\Delta$ is assumed to be small so that rejection events are mutually exclusive (we will take $\Delta \rightarrow 0$).
Hence,
\begin{equation}
	P_\mathrm{reject} = \sum_{j \neq a} P(j \ \mathrm{rejects} \ a).
\end{equation}

For the process to preserve the target measure, it must satisfy global balance:
\begin{equation}
	\sum_S \pi(S) P(S \rightarrow S') = \pi(S').
\end{equation}
Hence we must enumerate all the possible ways state $S$ can transition to $S'$.
\begin{equation}
	\begin{aligned}
		(\{x_i\}, a)
		&\;\xrightarrow{\text{accept}}\;
		(\{x_i'\}, a) \\
		(\{x_i\}, a)
		&\;\xrightarrow{\text{rejected by } b}\;
		(\{x_i\}, b)
	\end{aligned}
\end{equation}
Where $x_i' = x_i + \delta_{ia} \Delta x$.
It is then sufficient to require
\begin{equation}
	\begin{aligned}
		&\pi(\vec{x},i)P(\vec{x}',i\rightarrow\vec{x},i)\\
		&\quad+\sum_j\pi(\vec{x},j)
		P(\vec{x},j,v\rightarrow\vec{x},i)
		\overset{!}{=}\pi(\vec{x},i).
	\end{aligned}
\end{equation}
where $\vec{x}$ denotes the $N$ positions $(x_1, \cdots, x_N)$.

The factorized Metropolis accept probability is
\begin{equation}
	\begin{aligned}
		&P(\vec{x}', i \rightarrow \vec{x}, i) \\
		&\ = \prod_k \min (1, e^{-\beta (U(x_i' - x_k) - U(x_i - x_k))) } ) \\
		&\ \approx \prod_k \min (1, e^{-\beta U'(x_i - x_k) \Delta x } ) \\
		&\ = \prod_k \exp( - \beta \max( 0, U'(x_i - x_k) ) \Delta x) \\
		&\  \approx 1 - \sum_k \beta \max(0, U'(x_i - x_k)  \Delta x).
	\end{aligned}
\end{equation}
The approximation will become exact once $\Delta x \rightarrow 0$.

The individual rejection probability is 
\begin{equation}
	P(\vec{x},j\rightarrow\vec{x},i)
	=\beta\max(0,U'(x_j-x_i)\Delta x)+\mathcal O(\Delta x^2).
\end{equation}

Observe that
\begin{equation}
	\frac{\pi( \{ x_i' \} )}{\pi( \{ x_i \} )} \approx 1 + \beta \sum_j U'(x_i -x_j) \Delta x.
\end{equation}

Plugging in, the global balance condition becomes
\begin{equation}
	\begin{aligned}
		&\left(1+\beta\sum_j U'(x_i-x_j)\Delta x\right)\\
		&\quad\times\left(1-\beta\sum_j
		\max(0,U'(x_i-x_j)\Delta x)\right)\\
		&\quad+\beta\sum_j\max(0,U'(x_j-x_i)\Delta x)
		\overset{!}{=}1.
	\end{aligned}
\end{equation}
Expanding:
\begin{equation}
	\begin{aligned}
		&1 - \beta \sum_j \max(0, -U'(x_i - x_j) \Delta x) \\
		&\ + \beta \sum_j \max(0, U'(x_j - x_i) \Delta x) \overset{!}{\approx} 1.
	\end{aligned}
\end{equation}
Since $U'(x_j - x_i) = - U'(x_i - x_j)$, the condition holds, and hence the process satisfies global balance.

Taking $i$ to be the moving particle, the factorized acceptance probability at each step is
\begin{equation}
	\begin{aligned}
		&P_\mathrm{accept} = \prod_{j \neq i} \min ( 1 , e^{-\beta \Delta U_{ij}} ) \\
		& \ = \prod_{j \neq i } \exp( -\beta \max(0, \Delta U_{ij} ) ),
	\end{aligned}
\end{equation}
where $\Delta U_{ij} = U(x_i + t\Delta x - x_j) - U(x_i - x_j)$.
We hereby define the always increasing potential $V_{ij}$, which increases when $U(x_i - x_j)$ increases, but never decreases.
\begin{equation}
	\Delta V_{ij} \equiv \Delta V(x_i - x_j) = \max(0, \Delta U_{ij} ).
\end{equation}
\begin{equation}
	P_{\mathrm{accept}} = \prod_{j \neq i } \exp( -\beta \Delta V_{ij} )
\end{equation}

The probability of being rejected at step $T$ by particle $j$ is
\begin{equation}
	\begin{aligned}
		&P(\text{$j$ rejects at step $T$}) \\
		&\quad=\left(\prod_{t=1}^{T-1}P_\mathrm{accept}(t)\right)
		\left[1-P(\text{$j$ accepts at step $T$})\right] \\
		&\approx \exp \left( -\beta \sum_{t=1}^{T-1} \sum_{j \neq i} \Delta V_{ij}(t) \right)  \left( \beta \Delta V_{ij}(T) \right) \\
		&= \exp \left( -\beta \sum_{j \neq i} ( V_{ij}(T) - V_{ij}(0) ) \right)  \left( \beta  \Delta V_{ij}(T) \right) \\
		&\approx \exp \left( -\beta \sum_{j \neq i} ( E^{\mathrm{up}}_{ij}(T) ) \right)  \left( \beta \ dE^{\mathrm{up}}_{ij}(T)  \ \Delta \right)
	\end{aligned}
\end{equation}
where we have defined the accumulated pairwise upwards potential $E^{\mathrm{up}}_{ij}(T)$ to be $\sum\limits_{t=0}^T \Delta V_{ij}(t)$.
The rejection rule is equivalent to sampling independent exponential energy thresholds for the different $E_{ij}^{\mathrm{up}}$~\cite{PetersDeWith2012}.
Since at each infinitesimal step only one rejection is possible, we may simply sample the exponential distribution for each particle, and find which rejects first.
More precisely, each energy increase is sampled from the exponential:
\begin{align}
	&E_{ij}^{\mathrm{up}} = -\frac{1}{\beta} \log(\mathrm{ran}(0,1)). 
\end{align}
This can then be inverted to find $T$ at which particle $j$ rejects $i$.
Sampling for the $N$ different $j$'s, we find the shortest $T_\mathrm{event}$, and advance towards it in simulation.
$j$ then becomes the active particle.
As it stands, the algorithm would only sample at events, which introduces a bias. Concretely, for a simulation of hard spheres, all samples would be configurations with one pair in contact: not a generic configuration from the distribution.
To avoid this event-conditioned sampling bias, configurations are recorded at fixed time intervals $dt$, irrespective of when events occur~\cite{Krauth2021ECMC}.
Repeating this process retains the target distribution of the lifted
discrete-step Monte Carlo algorithm while preserving its directed updates.

We now want to consider the fracton Hamiltonian, which involves both positions and momenta.
We extend the ECMC algorithm.
First, for some number of steps, we sample with momenta frozen, using a Markov kernel that preserves $P(\{x_i\}\mid\{p_i\})$.
The algorithm then switches to momenta and, with positions frozen, applies a kernel that preserves $P(\{p_i\}\mid\{x_i\})$.
Each conditional update preserves the joint target because the frozen variables keep their marginal weight while the updated variables retain the corresponding conditional distribution. 
Their composition therefore preserves $P(\{x_i,p_i\})$; irreducibility is still required for convergence~\cite{Tierney1994}.
In the main text (\cref{sec:ecmc-algorithm}), we quantitative match numerical results from the ECMC construction and the naive single-step MC algorithm, which is trivially irreducible.
With this in mind, we assume ECMC is irreducible for the kernels studied.
	
\bibliography{references}
	
\end{document}